\documentclass[10pt,prb,aps,showpacs,twocolumn,unsortedaddress]{revtex4-1}
\usepackage{graphicx,bm}
\usepackage{amsmath}
\usepackage{amssymb}
\usepackage{epsfig}
\usepackage{epsf}
\usepackage{verbatim}
\usepackage{hyperref}
\usepackage[usenames]{color}

\newcommand{\beqa}{\begin{eqnarray}}
\newcommand{\eeqa}{\end{eqnarray}}
\newcommand{\beq}{\begin{equation}}
\newcommand{\eeq}{\end{equation}}

\def\bra#1{\mathinner{\langle{#1}|}} 
\def\ket#1{\mathinner{|{#1}\rangle}}

\newcommand{\mbf}[1]{\mbox{\boldmath$#1$}}
\newcommand{\Js}{\mathbf{J}^{s}}

\newcommand{\bs}[1]{\boldsymbol{#1}}

\begin{document}
%\title{Piezospintronic effect in antiferromagnetic honeycomb lattice}
\title{Piezospintronic effect in honeycomb antiferromagnets}
\author{Camilo Ulloa$^{1,4}$}
\email{C.Ulloa@uu.nl}
 \author{Roberto E. Troncoso$^{2,3}$}
\author{Scott A. Bender$^{4}$}
\author{R. A. Duine$^{4,5}$}
\author{A. S. Nunez$^1$}
\affiliation{${}^1$Departamento de F\'isica, Facultad de Ciencias F\'isicas y 
Matem\'aticas, Universidad de Chile, Casilla 487-3, Santiago, Chile}
\affiliation{${}^2$Department of Physics, Norwegian University of Science and Technology, NO-7491 Trondheim, Norway}
\affiliation{${}^3$Departamento de F\'isica, Universidad T\'ecnica Federico Santa Mar\'ia, Avenida Espa\~na 1680, Valpara\'iso, Chile}
\affiliation{${}^4$Institute for Theoretical Physics, Utrecht University,
Princetonplein 5, 3584 CC Utrecht, The Netherlands}
\affiliation{${}^5$Department of Applied Physics, Eindhoven University of Technology,
P.O. Box 513, 5600 MB Eindhoven, The Netherlands}

\begin{abstract}
The emission of pure spin currents by mechanical deformations, the piezospintronic effect, in antiferromagnets is studied. We characterize the piezospintronic effect in an antiferromagnetic honeycomb monolayer in response to external strains. It is shown that the strain tensor components can be evaluated in terms of the spin Berry phase. In addition, we propose an experimental setup to detect the piezospin current generated in the piezospintronic material through the inverse spin Hall effect. Our results apply to a wide family of two-dimensional antiferromagnetic materials without inversion symmetry, such as the transition-metal chalcogenophosphates materials MPX$_3$ (M=V, Mn; X=S, Se, Te) and NiPSe$_3$.
\end{abstract}

%Spin polarized transport, 72.25.-b
%Spin transport (magnetoelectronics), (spin transport effects )75.76.+j 
%Antiferromagnetics, 75.50.Ee

\maketitle
%++++++++++++++++++++++++++++++++++++++++++++++++++++++++++++

%{\color{blue}
%\begin{itemize}
%	\item Time reversal versus Spin reversal.
%	\item Unit cell in the Fig. \ref{fig:model}.
%	\item Needed of the back spin current and last sentence of sec. IV.
%	\item Outlooks
%\end{itemize}	
%}

%%%%%%%%%%%%%%%%%%%%%%%%%%%%%%%%%%%%%%%%%%%%%%%%%%%
\section{Introduction} 
%%%%%%%%%%%%%%%%%%%%%%%%%%%%%%%%%%%%%%%%%%%%%%%%%%%
Spintronics is one of the most promising areas in condensed matter from the point of view of development of novel devices that can enhance or directly replace conventional electronics\cite{Sinova,Duine}. This has motivated intense studies to understand the mutual relation between spin currents and magnetic properties\cite{Wolf}. In this context, antiferromagnets (AFs) have recently gained attention due to their favorable properties\cite{NunezAFM} and abundance in nature\cite{Antiferromagnets}. Compared with conventional ferromagnets, AFs lack macroscopic magnetization\cite{landau} and furthermore can be operative at much higher frequencies than ferromagnets\cite{AF_Terahertz}. The absence of stray fields makes them robust against perturbation due to magnetic fields. Moreover, AFs have also opened a new branch in spintronics by hosting topological matter, such as Weyl semimetals\cite{} and topological insulators\cite{}. 

AFs can also be the cornerstone of spin-current generation. It has been shown that spin angular momentum can be transported through AF$|$NM (normal metal) heterostructures, in the form of pumped spin and staggered spin currents\cite{RanCheng}, associated with the dynamics of the magnetization and staggered field (N\'eel order) respectively. An alternative route for the generation of spin currents has recently been proposed, which is based on the coupling between mechanical distortions and spin degrees of freedom, namely the piezospintronic effect\cite{Alvaro}. Unlike the related piezoelectric \cite{piezoelectric} and piezomagnetic effects \cite{piezomagnetic}, this phenomena is restricted to appear in systems with the concomitance of time reversal($\cal T$) and inversion ($\cal I$)  symmetry breaking. Although in principle, a crystal might display simultaneously the piezoelectric, piezomagnetic and piezospintronic effects.

From a phenomenological point of view, a magnetic crystal under mechanical deformations gives rise in linear response to a spin dipolar moment, $P^s_{\sigma;j}=\sum_{kl}\lambda_{\sigma;jkl} {u}_{kl}$, with $\lambda$ the piezospintronic pseudo-tensor\cite{Nye}, where $\sigma$ and $j$ label spin and position components respectively, and ${u}_{kl}=\left(\partial_lu_k+\partial_ku_l\right)/2$ the strain tensor\cite{LANDAU} with ${\vec u}$ the deformation field (sketched in Fig.\ref{fig:model}). Under inversion $\lambda$ change sign and therefore like the piezoelectric\cite{piezoelectric} and piezomagnetic effects\cite{piezomagnetic}, the piezospintronic effect is restricted only to crystals lacking a center of inversion. Similarly, the spin dipole moment is odd under time reversal, so a system with a non-vanishing piezospintronic tensor must have broken time reversal invariance. 

The spin dipolar moment and spin currents are linked\cite{Alvaro}, through the standard definition \cite{QNiu} of spin currents, by 
\begin{align}\label{eq:piezospincurrent}
{J}^s_{\sigma,j}=\frac{{\rm d} {P}^s_{\sigma;j}}{{\rm d}t}.
\end{align}
Is intuitive to realize that crystals classes invariant under $\mathcal{T}\mathcal{I}$ will respond with a pure spin-current to an external strain, i.e., displaying exclusively the piezospintronic effect without giving rise to charge currents. The simple way to understand that is to require $\cal T-$ and $\cal I-$symmetry breaking and thus, each spin component manifests opposite piezoelectric effects \cite{Nye,LANDAU}. Under inversion the direction of each piezoelectric effect is reversed, while the spin labels remain unchanged and thus there is a reversal of the piezospin current. An additional spin reversal, through the action of ${\cal T}$, will restore the original current. Therefore, it is expected that crystals classes invariant simultaneously under spin reversal and spatial inversion will respond with a pure spin-current to an external deformation. Some AFs structures, like antiferromagnetic honeycombs, represent natural systems to explore this effect since they bring together alternating spin configurations that additionally break inversion symmetry.

In this work we present detailed calculations concerning the piezospintronic tensor of an antiferromagnetic honeycomb lattice. These calculations were performed within the tight-binding approximation. In heterostructures as AF$|$NM (normal metal) this effect can be tested via inverse spin Hall effect (ISHE) measurements\cite{Saitoh}.
 
This paper is organized as follows. In Sec. \ref{sec3}, we compute the piezospintronic properties of an AF honeycomb lattice. In Sec. \ref{sec4} we propose an experimental setup to measure the piezospintronic spin current generated at the interface with a normal metal (NM) through the ISHE in the NM. Finally, we finish in Sec. \ref{sec5} with the conclusions and discussions.

%%%%%%%%%%%%%%%%%%%%%%%%%%%%%%%%%%%%%%%%%%%%%%%%%%%
\section{Antiferromagnetic honeycomb}\label{sec3}
%%%%%%%%%%%%%%%%%%%%%%%%%%%%%%%%%%%%%%%%%%%%%%%%%%%
\subsection{Model}
\begin{figure}[htbp]
   \centering
   \includegraphics[width=0.5\textwidth]{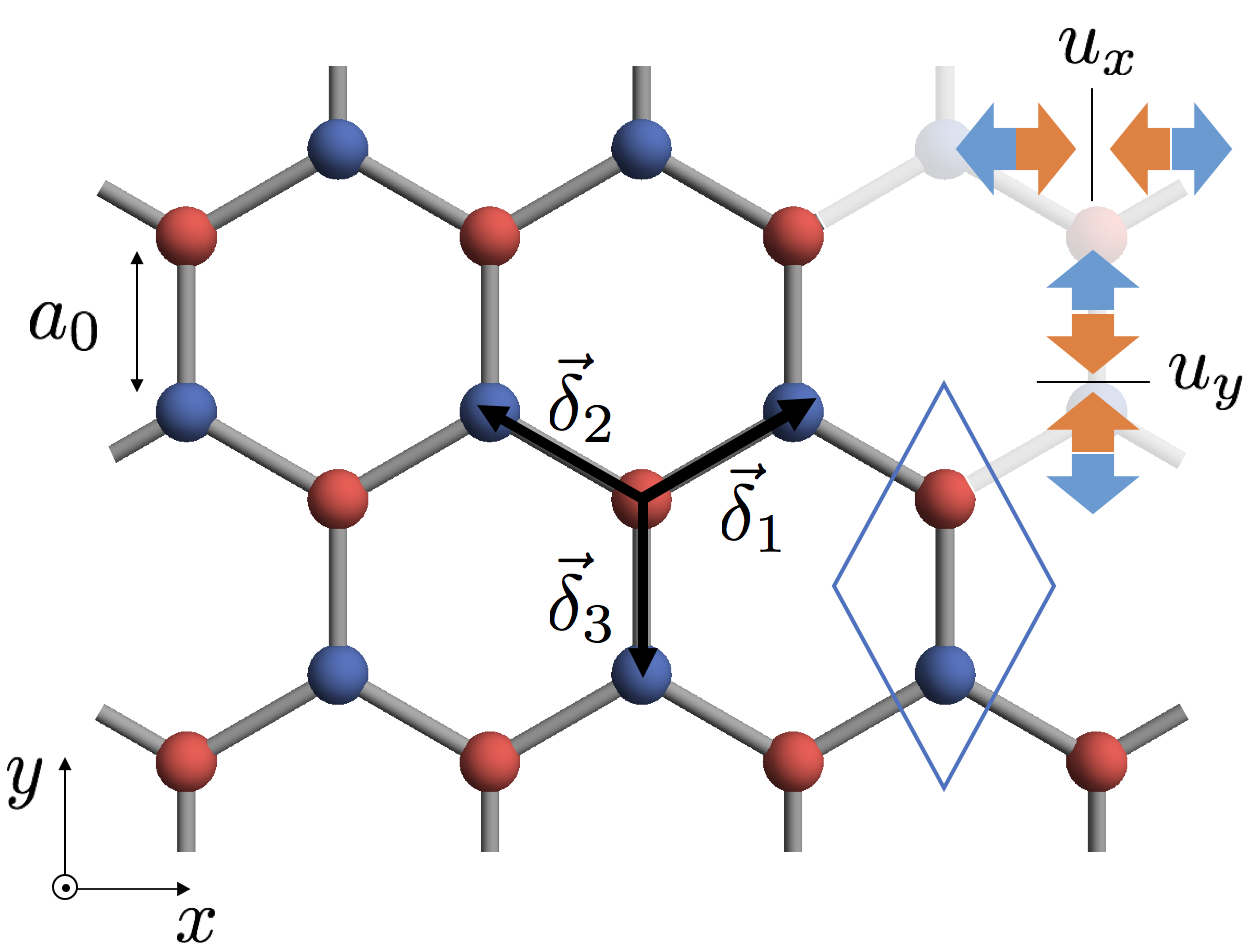} 
   \caption{Antiferromagnetic honeycomb lattice in the $xy$ plane. Red and blue spheres represent the different spin species, spin up and down respectively, whose spin polarizations are along $z$-direction. Note that this structure lacks $\mathcal{T}$ and $\mathcal{I}$ symmetries but is invariant under the $\mathcal{TI}$ transformation, thus fulfilling the requirements to exhibit pure piezospintronic response. In the tight-binding approximation the first nearest neighbours of each site are described by the vectors $\vec{\delta}_i$ with $a_0$ the unperturbed lattice constant. Strains along $x-$ and $y-$directions are schematically depicted and represented by the strain deformation field $u_{x}$ and $u_{y}$, respectively.}
   \label{fig:model}
\end{figure}

We consider a honeycomb lattice with staggered spin array lying in the $xy$-plane, as is described in Fig. \ref{fig:model}. The antiparallel lattices of spins, represented by the red and blue dots in Fig. \ref{fig:model}, are oriented along the $z$-axis with spin polarization ${\mathbf\Omega}^{\text{A,B}}_i=\pm\Delta \mathbf{z}$. Under spatial inversion around the center of the unit cell, represented by the rhombus in Fig. \ref{fig:model}, the position of both sub lattices is reversed, i.e., blue and red dots are interchanged. A subsequent time reversal operation flips the local spin and thus reverts the effect of spatial inversion. The system is invariant under $\mathcal{T}\mathcal{I}$ and therefore we expect it to display a pure piezospintronic effect. 
%%%%%%%%%%FIGURE1
%%%%%%%%%%
Complementing the local exchange term in our model we also consider hopping to nearest neighbours. The net Hamiltonian is
\begin{equation}\label{hamiltonian}
\mathcal{H}=-\sum_{\langle {\boldsymbol{i}},{\boldsymbol j}\rangle, \sigma}\text{t}_{{\boldsymbol i}{\boldsymbol j}}\left(c^\dagger_{{\boldsymbol i}\sigma}c^{\phantom\dagger}_{{\boldsymbol j}\sigma}+ h.c. \right)+\Delta\sum_{{\boldsymbol i};\sigma,\sigma^\prime}\eta_{{\boldsymbol i}}c^\dagger_{{\boldsymbol i}\sigma}\sigma^z_{\sigma\sigma^\prime}c^{\phantom\dagger}_{{\boldsymbol i}\sigma^\prime}
\end{equation}
where $c^\dagger_{{\boldsymbol i}\sigma}(c_{{\boldsymbol i}\sigma})$ is the operator that creates(annihilates) an electron with spin $\sigma$ on site ${\boldsymbol i}$, and $\eta_{\boldsymbol{i}}=\pm1$ depending on the sublattice. The hopping matrix element is $\text{t}_{{\boldsymbol i}{\boldsymbol j}}$, the energy difference between the spin species is $\Delta$, and $\sigma^z$ is the $z-$component of the Pauli matrix vector. The Hamiltonian in Eq. (\ref{hamiltonian}) is Fourier transformed to momentum space and written as 
\begin{align*}
\mathcal{H}_a=\left(   \begin{matrix} 
      \Delta & \gamma_k \\
      \gamma^*_k & -\Delta \\
   \end{matrix}
\right), &\;\;\;
\mathcal{H}_b=\left(   \begin{matrix} 
      -\Delta & \gamma_k \\
      \gamma^*_k & \Delta \\
   \end{matrix}
\right),
\end{align*}
\begin{equation*}
\mathcal{H} = \left(   \begin{matrix} 
      \mathcal{H}_a & 0 \\
      0 & \mathcal{H}_b \\
   \end{matrix}
\right);
\end{equation*}
with $\gamma_k=\sum_{\boldsymbol i} \text{t}_{{\boldsymbol i},{\boldsymbol i}+\vec{\delta}_{{\boldsymbol i}}} 	\exp(i \vec{k}\cdot \vec{\delta}_i)$. In the follwing we label $\text{t}_{j}$ the hopping amplitude conecting a site with his neighbour $\vec{\delta}_j$ (see Fig. \ref{fig:model}). At this point we can draw an analogy between the model Hamiltonian we are proposing and the tight-binding model of Boron-Nitride (BN) monolayers\cite{BN_Droth}. Sharing the honeycomb structure we see how our model reduces to the BN for each spin species, but with an opposite role for each sub lattice. The effect we are looking for follows from the piezoelectric response of BN and will share all the symmetry properties with it.

%%%%%%%%%%%%%%%%%%%%%%%%%%%%%%%%%%
\subsection{Piezospintronic tensor of a honeycomb antiferromagnet}
%%%%%%%%%%%%%%%%%%%%%%%%%%%%%%%%%%
As the crystal belongs to the point group $\bar{6}m2\ (D3h)$\cite{Nye}, following the symmetry analysis of BN we can conclude the following property of the piezospintronic tensor\cite{BN_Droth}. All the components of the tensor are zero except for
\begin{align*}
\lambda_{z;yyy}=-\lambda_{z;yxx}=-\lambda_{z;xyx}.
\end{align*}
Our task is then reduced to the evaluation of only one of the components of the tensor, e.g. $\lambda_{z;yyy}$. We evaluate the net spin dipolar moment created by a deformation of the lattice along the $y-$direction. With the deformation the different hopping amplitudes will change, a simple geometrical analysis leads to
\begin{align*}
d\text{t}_1=d\text{t}_2=\frac{1}{2}d\text{t}_3=\left(\frac{\partial\text{t}}{\partial a}\right)du_{yy}
\end{align*}
where $\text{t}$ is the hopping amplitude at an inter-atomic distance $a$. The net spin dipolar moment generated is given by\cite{resta}
\begin{align*}
{\rm d}{\bf P}^S_{z;y}={\bf A}^{\text{t}_1}_{z,y}{\rm d}\text{t}_1+{\bf A}^{\text{t}_2}_{z,y}{\rm d}\text{t}_2+{\bf A}^{\text{t}_3}_{z,y}{\rm d}\text{t}_3,
\end{align*}
where ${\bf A}^{\text{t}_{\alpha}}_{z,y}$ is defined as
\begin{align*}
{\bf A}^{\text{t}_{\alpha}}_{z,y}\equiv\frac{\partial {\bf P}^s_{z;y}}{\partial \text{t}_{\alpha}},
\end{align*}
and can be evaluated in terms of spin Berry phases\cite{Alvaro, King1993} which depend on the electronic Bloch states $\ket{\phi_{\nu}}$ as
\begin{equation}\label{spin_berry}
{\bf A}^{\text{t}_{\alpha}}_{i,j}=-\sum_\nu\int_{BZ}\frac{{\rm d}^2k}{(2\pi)^2}n_\nu(k)\text{Im}
\left\langle \frac{\partial \phi_\nu}{\partial k_j}\right|\sigma_i\left|\frac{\partial \phi_\nu}{\partial \text{t}_{\alpha}}\right\rangle.
\end{equation}

The symmetry of the hexagonal lattice enforces a relation among the different $\mathbf{A}_{i,j}^{\text{t}_{\alpha}}$ that reads, ${\bf A}^{\text{t}_1}_{z,y}={\bf A}^{\text{t}_2}_{z,y}=-\frac{1}{2}{\bf A}^{\text{t}_3}_{z,y}$, leading to the final expression for the piezospintronic tensor,
\begin{equation}
\label{tensor_piezo} \lambda_{z;yyy}=-\frac{1}{2}\left(\frac{\partial \text{t}}{\partial a}\right){\bf A}^{\text{t}_3}_{z,y}.
\end{equation}

\begin{figure}[htbp] 
    \centering
    \includegraphics[width=0.5\textwidth]{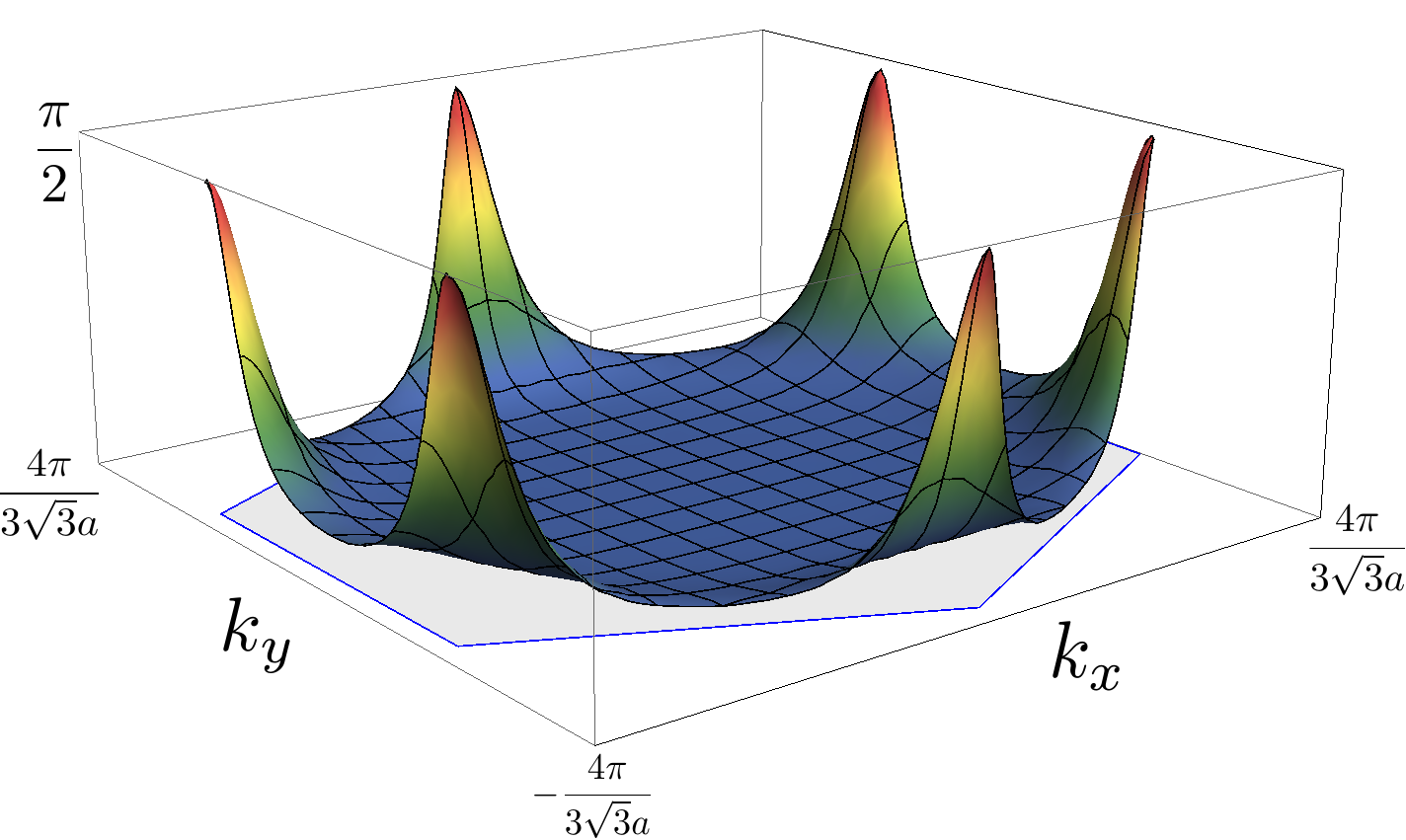} 
    \caption{Berry curvature in the first Brillouin zone of antiferromagnetic honeycomb (see Eq. \eqref{spin_berry}). The integral of this function leads us directly to the value of the piezospintronic tensor $\lambda_{z;yyy}$ (see Eq. \eqref{tensor_piezo}).  Due to the presence of the local energy $\Delta$ there is an asymmetry in the two sublattices, which opens a gap in the spectra around the Dirac points.}
    \label{fig:integrand}
 \end{figure}

\begin{figure}[htbp] %  figure placement: here, top, bottom, or page
   \centering
   \includegraphics[width=0.5\textwidth]{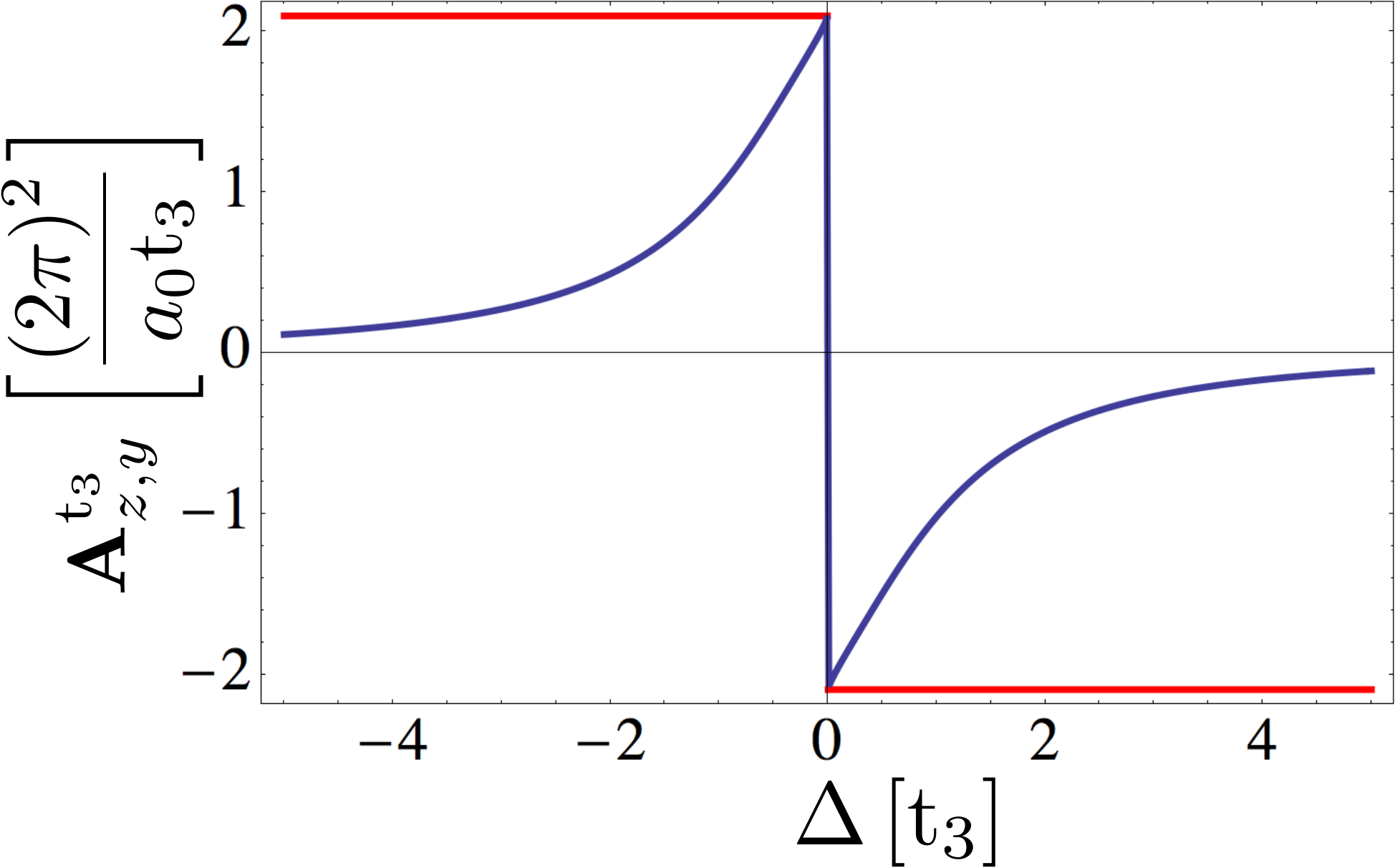} 
   \caption{Result of the integration of ${\bf A}^{\text{t}_3}_{z,y}$ for antiferromagnetic honeycomb as function of the local energy $\Delta$. The blue line corresponds to the numerical integration of the Berry curvature in the first Brillouin zone. The red line shows the exact calculation in the long wavelength limit. The maximum(minimum) value of the curve is $2\pi/3$($-2\pi/3$) as we expect from the long wavelength approximation (see Eq.\eqref{eq:A_long}). }
   \label{fig:tensor}
\end{figure}

The integrand in the expression for ${\bf A}^{\text{t}_3}_{z,y}$ is displayed in Fig. \ref{fig:integrand}. It displays well defined maxima around the corners of the Brillouin zone.
%%%%%%%%%%FIGURE2
 %%%%%%%%%%
The integral is complicated and needs to be evaluated numerically, which obtained result is displayed in Fig. \ref{fig:tensor}. 
Around one of the Dirac points of the Brillouin zone, however the Berry curvature can be approximated analytically, yielding 
\begin{equation}\label{eq:A_long}
\mathbf{A}^{\text{t}_3}_{z,y}=-\dfrac{\text{sign}(\Delta)}{6\pi a_0\text{t}}.
\end{equation} 
The independence of the magnitude of $\Delta$ in this result arises from the expression of the eigenstates $\ket{\phi_{\nu}}$ in the long wavelength limit. As his $j$-th component is proportional to $\text{t} k_{j}$ we can replace $\partial_{\text{t}_j}\rightarrow \partial_{k_j}$ obtaining an expression proportional to a Chern number\cite{TKNN}.

%%%%%%%%%%%%%%%%%%%%%%
\section{Detection of piezospin currents}\label{sec4}
%%%%%%%%%%%%%%%%%%%%%%
Now we propose an experimental setup to perform an indirect detection of the spin current generated by the piezospintronic effect. We consider two adjacent materials, as is shown in Fig. \ref{fig:setup}, one being piezospintronic and the other being a normal metal with strong spin-orbit coupling(SOC). As is well known, due to the SOC, a charge current ${\bf J}_q$ flowing in the NM converts into a pure spin current(SHE), and vice versa(ISHE)\cite{Saitoh}. Based on the above effect we expect to measure a charge Hall current as a result of the piezospin current induced at the interface. The process of injection of a spin current into a metal has been widely studied, as for example in Ref. [{\onlinecite{injection}]. We consider that a spin current is generated in the piezospintronic material with no loss of spin angular momentum in the bulk. Thus, the total spin current at the interface is $\Js_{\sigma}(y=0)={\bf J}_{s,\sigma}$. For simplicity a perfect transmission of spin current through the interface is assumed. Under this assumption we calculate analytically the charge current generated in the normal metal in terms of the spin current emitted from the piezospintronic material. 
%%%%%%%%%%FIGURE3
\begin{figure}[htbp] 
\centering
\includegraphics[width=0.47\textwidth]{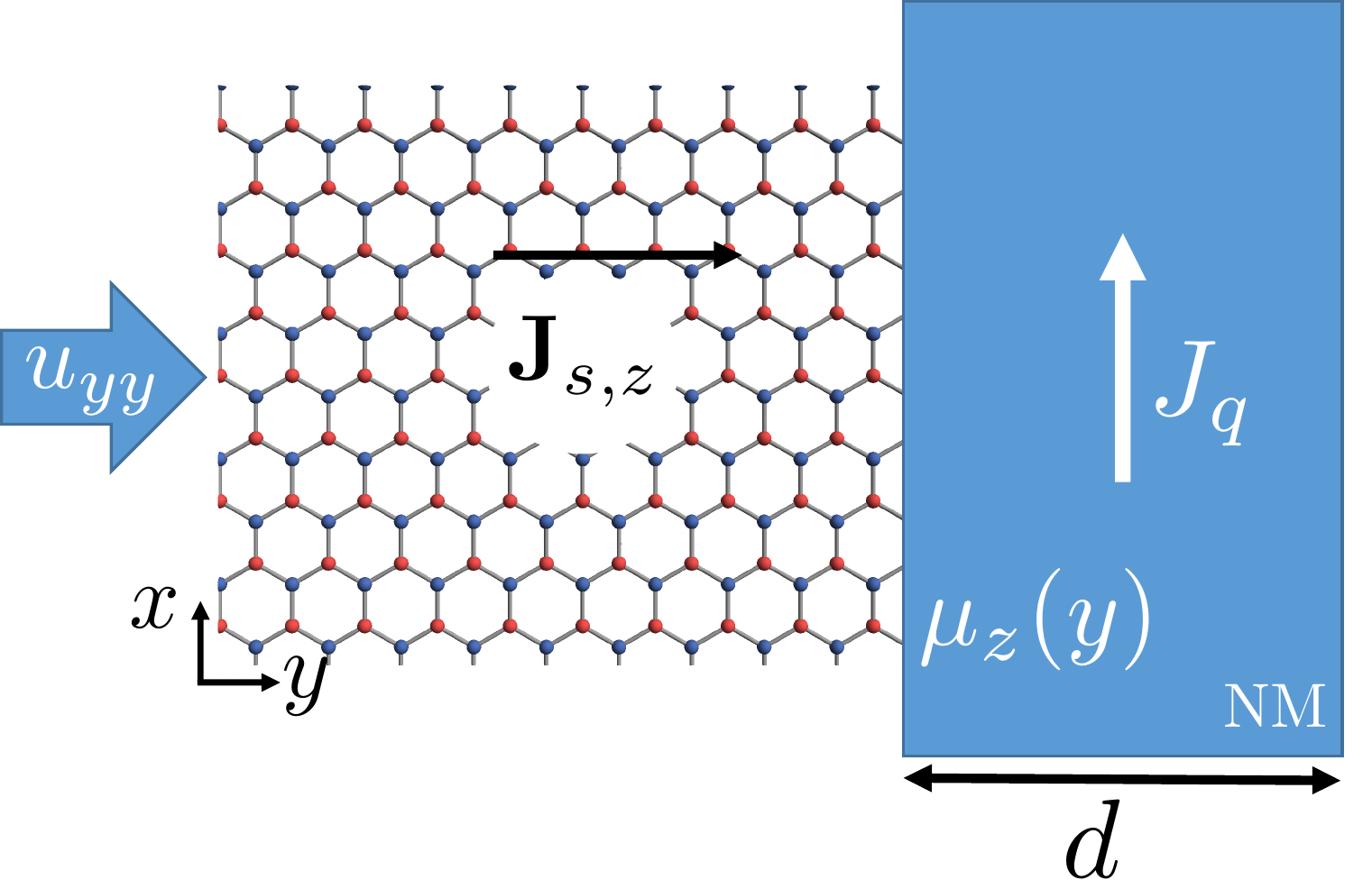}
\caption{Schematic setup for the detection of piezospin currents. The proposal is based on similar geometries as is used for the measurement of the spin Seebeck effect\cite{Saitoh2}. Under a strain $u_{yy}$ a spin current ${\bf J}_{s,z}={J}_{s,z}\hat{\bs y}$ is induced in the piezospintronic material and injected into the NM. In turn this results in a spin accumulation $\mu_{z}(y)$ on the NM (translation symmetry in $z$ is assumed). Due to the spin orbit coupling in the NM a transverse charge current $J_q$ along $x-$direction, i.e., a ISHE signal.}
\label{fig:setup}
\end{figure}
%%%%%%%%%%

To analyze the connection between the spin and charge currents in the metallic material we solve the spin diffusion equation for the spin accumulation\cite{spin_diff} $\boldsymbol{\mu}_s(\mbf{x})$, 
\begin{align}\label{eq:diffEq}
\nabla^2 \boldsymbol{\mu}_s=\frac{\boldsymbol{\mu}_s}{\ell^2_s},
\end{align} 
where $\ell_s$ stands for the characteristic spin diffusion length of the NM. The boundary conditions for Eq. (\ref{eq:diffEq}) enforce continuity for the spin current which reads,
\begin{align}
\partial_y\boldsymbol{\mu}_s\left.\right|_{y=0}&=-\frac{G_0}{\sigma}{\bf J}^{\text{net}}_{s,z}\\
\partial_y\boldsymbol{\mu}_s\left.\right|_{y=d}&=0,
\end{align}
where $G_0=2e^2/h$ is the quantum of conductance, $\sigma$ and $d$ are the conductivity and the thickness of the NM. ${\bf J}^{\text{net}}_{s,z}$ is the net spin current flowing through into the NM. The net spin current is the sum of the injected piezospin current, as given in Eq. (\ref{eq:piezospincurrent}), and a backflow spin current ${\bf J}^{\text{back}}_{s}$ in the opposite direction due to the induced spin accumulation on the NM side of the interface. In the calculation of the spin Hall current we disregard spin transfer torques generated by the spin current in the normal metal acting on the antiferromagnet.
%For simplicity, it is assumed that the NM is an ideal spin sink interface, i.e., the injected spin current is fully absorbed, and thus we disregard the spin current back flow. 
Moreover, without loss of generality the piezospin current is considered to flow in the $y-$direction and polarized along the $z-$ axis. Additionally, in the bulk of NM spin and charge currents are related through the relations 
\begin{align}
{\bf J}_q&\label{eq:chargecurrent}=\dfrac{\sigma}{e}{\boldsymbol\nabla}\mu-\dfrac{\sigma'}{2e}{\boldsymbol\nabla}\times{\boldsymbol\mu}_s,\\
\dfrac{2e}{\hbar}{\bf J}_{s,z}&\label{eq:spincurrent}= -\dfrac{\sigma}{2e}\partial_{y}{\boldsymbol\mu}_s-\dfrac{\sigma'}{e}{\bf z}\times{\boldsymbol\nabla}\mu
\end{align}
%\begin{align}
%J_i&=\dfrac{\sigma}{e}\partial_{i}\mu-\dfrac{\sigma'}{2e}\epsilon_{ijk}\partial_j\mu_k,\\
%\dfrac{2e}{\hbar}J_{ji}^{S}&= -\dfrac{\sigma}{2e}\partial_{i}\mu_{j}-\dfrac{\sigma'}{e}\epsilon_{ijk}\partial_{k}\mu
%\end{align}
with $\mu$ the electronic chemical potential and $\sigma'$ the spin Hall conductivity in the NM\cite{spin_hall}. Solving Eqs. (\ref{eq:diffEq}$-$\ref{eq:spincurrent}) leads to an induced Hall charge current density along the $x-$direction
\begin{equation*}
%\left<J_x\right>=2\pi G_0\dfrac{\ell_s}{d}\tan\theta_{H} \tanh\left ( \dfrac{d}{2\ell_s}\right )\lambda_{z;ykl}\dfrac{{\rm d\mathbf{u}}_{kl}}{{\rm d}t},
\left<J_q^{x}\right>=\Gamma\sum_{kl}\lambda_{z;ykl}\dfrac{{\rm d{u}}_{kl}}{{\rm d}t},
\end{equation*}
where $\Gamma= \dfrac{2e}{\hbar}\dfrac{\ell_s}{d}\tan\theta_{H} \tanh\left ( \dfrac{d}{2\ell_s}\right )$, the Hall angle is $\theta_H=\arctan\left[\sigma'/\sigma\right]$, and $\langle ...\rangle$ denotes a thickness average.
%$\Gamma= \dfrac{2e}{\hbar}\dfrac{\ell_s}{d}\tan\theta_{H} \tanh\left ( \dfrac{d}{2\ell_s}\right )$
This effect can be measured by making use of materials with huge potential for spintronic devices\cite{Li2016}, for example the transition metal chalcogenophosphates MPX$_3$ (M=V, Mn; X=S, Se, Te) and NiPSe$_3$. These materials are 2D semiconductors in which the transition-metal atoms of the compound are organized in a honeycomb lattice. Recent theoretical studies\cite{chittari,lee} have shown that these materials might exhibit a N\'eel order in the ground state which is not affected under strain. Nevertheless this setup also works with materials without $\mathcal{TI}$ symmetry. In that case there will be an additional piezoelectric response, but the charge current generated in that process will generate a transversal signal in the metal which will not affect the Hall signal.

It is worth commenting that a reciprocal effect is also expected. 
%{\color{red}The thermodynamic conjugate of the spin current, a spin force, was identified by Shi et al\cite{QNiu}. The physical relationship between the spin current and the spin force allow us to state that a converse piezospintronic effect is manifested in system that have the direct piezospintronic response}. 
From Onsager's relations\cite{Onsager,Landau-stat} a stress is expected  in response to a spin-current injected into the system
\begin{align}
\label{eq: piezo-spintronic crystal reciprocal}
{\mathbf{s}}_{ij}=\sum_{lm}\tilde{\lambda}_{l;mij} \Js_{l;m},
\end{align}
where ${\mathbf{s}}$ stands as the stress tensor\cite{LANDAU} appearing in response to the spin current $\Js$.  This converse piezospintronic effect might lead to novel mechanisms to detect pure spin currents. In fact, this effect might be useful is the mechanical resonance of the piezospintronic material. The idea is to consider an AF-FM interface with the FM under ferromagnetic resonance. This will inject a spin current in the piezospintronic and due to the reciprocal piezospintronic effect (see Eq. \eqref{eq: piezo-spintronic crystal reciprocal}). With the proper excitation frequency it might get even into a resonant state.

%%%%%%%%%%%%%%%%%
\section{Conclusions}\label{sec5}
%%%%%%%%%%%%%%%%%
In this paper we have discussed the possibility of generating and detecting pure spin currents via the piezospintronic effect in honeycomb antiferromagnets. We discussed the principal characteristics of this effect and the symmetry properties that lead to a pure spin current in response to strain. We characterized the piezospintronic response of a honeycomb antiferromagnetic layer, which fulfills the symmetry conditions to develop a pure piezospintronic response, and calculated its piezospintronic tensor. In the long wavelength approximation we showed that the relevant coefficients of the piezospintronic tensor are proportional to a Chern number. Finally, we proposed an experimental setup to measure the spin current generated in this way by converting it into an electric current through the ISHE. This work extends the grounds for spin-mechanics\cite{spinmec} systems because it provides a direct coupling between spin current and strain.
%%%%%%%%%%%%%%%%
\section{Acknowledgements} 
%%%%%%%%%%%%%%%%
It is a pleasure to thank J. Fern\'andez-Rossier for fruitful conversations. ASN and CU would like to thank funding from grants Fondecyt Regular 1150072. ASN also acknowledges support from Financiamiento Basal para Centros Cient\'ificos y Tecnol\'ogicos de Excelencia, under Project No. FB 0807 (Chile). RET thanks funding from  Fondecyt Postdoctorado 3150372 and Fondecyt Regular 1161403. RD is member of the D-ITP consortium, a program of the Netherlands Organisation for Scientific Research (NWO) that is funded by the Dutch Ministry of Education, Culture and Science (OCW). This work is in part funded by the Stichting voor Fundamenteel Onderzoek der Materie (FOM).

\appendix
\section{Calculation of the Berry phase}
To calculate the Berry phase we use the coherent states 
$$\ket{\mathbf{n}}=\left (\begin{array}{c}
e^{i\varphi}\cos\theta/2 \\ 
\sin\theta/2
\end{array} \right ), $$
along the direction of the vector
\begin{equation}\label{eq:n}
\mathbf{n}=\left (\sum_{i}\text{t}_i\cos(\mbf{k}\cdot\vec{\delta}_{i}),\sum_{i}\text{t}_i\sin(\mbf{k}\cdot\vec{\delta}_{i}),\Delta\right ),
\end{equation}
where $\theta$ and $\varphi$ are the polar and azimuthal angles respectively in spherical coordinates. 
In this representation Eq.\eqref{spin_berry} can be written as
$$\mathbf{A}^{\text{t}}_{z,j}=\int_{BZ} \dfrac{{\rm d}^{2}k}{(2\pi)^{2}}\ \text{Im}\bra{\nabla_{Q}\mathbf{n}}\times\ket{\nabla_{Q}\mathbf{n}},$$
where $Q=\text{t}, k_j$. Hence, we can calculate the relevant contribution 
\begin{equation}\label{eq:BP}
{\bf A}^{t_3}_{z,y}=\int_{BZ}\frac{{\rm d}^2k}{(2\pi)^2}
\dfrac{\sin\theta}{4}\left ( \dfrac{\partial\theta}{\partial\text{t}_3}\dfrac{\partial\varphi}{\partial k_y}-\dfrac{\partial\theta}{\partial k_y}\dfrac{\partial\varphi}{\partial\text{t}_3}\right).
\end{equation}
The numerical solution of this integral is shown in the blue line of Fig.\ref{fig:tensor}. 

In the long wavelength limit we perform an expansion of the eigenstates in $k$ around the Dirac point $K_+=(4\pi/3\sqrt{3}a_0,0)$. Around this point Eq.\eqref{eq:n} becomes
\begin{equation*}
\mathbf{n}_{K_+}=\left (- \dfrac{3}{2}a_0\text{t} k_{x},-\dfrac{3}{2}a_0\text{t}k_{y},\Delta\right ),
\end{equation*}
which, following Eq.\eqref{eq:BP}, lead us to the expression
$$\mathbf{A}_{z,y}^{\text{t}_3} = -2\int\dfrac{{\rm d}^{2}k}{(2\pi)^{2}}\dfrac{3a_0\text{t}_3\Delta }{(9a_0^{2}\text{t}_3^{2}(k_x^{2}+k_y^{2})+4\Delta^2)^{3/2}}.$$
The result of this integral is exactly Eq.\eqref{eq:A_long} and is shown in the red line of Fig.\ref{fig:tensor}.

%We start diagonalizing the Hamiltonian of Eq.\eqref{hamiltonian} in the coherent state basis $\ket{\mathbf{n}}$, which corresponds to the unit vector along the direction of $\left (\Sigma_i \text{t}_i \cos(\mathbf{k}\cdot\vec{\delta}_i) ,\Sigma_i \text{t}_i \sin(\mathbf{k}\cdot\vec{\delta}_i) ,\Delta\right ).$  To calculate the relevant contribution of Eq.\eqref{spin_berry} we perform an expansion in $k$ around the Dirac point $K_+=(4\pi/3\sqrt{3}a_0,0)$. Close to $K_+$ the eigenstates become $\ket{\mathbf{n}_{K_+}}$ which is the coherent state in the direction of  $(-3\text{t}a_0k_x/2,-3\text{t}a_0k_y/2,\Delta)$. With this ingredients the berry curvature with be
%$$ \text{Im}\bra{\partial_Q \mathbf{n}}\times\ket{\partial_{Q}\mathbf{n}},$$

%Eq.\eqref{spin_berry} becomes 
%$$\mathbf{A}_{z,y}^{\text{t}_3} = -2\int\dfrac{d^{2}k}{(2\pi)^{2}}\dfrac{3a_0\text{t}_3\Delta }{(9a_0^{2}\text{t}_3^{2}(k_x^{2}+k_y^{2})+4\Delta^2)^{3/2}}.$$
%The result of this integral is exactly Eq.\eqref{eq:A_long}.
\bibliographystyle{elsarticle-num}

\end{document}